\documentclass[
  reprint,
  superscriptaddress,
  amsmath,amssymb,
  aps,prx,
  superscript,
  citeautoscript
]{revtex4-2}

\usepackage{graphicx} % Required for inserting images
\usepackage{booktabs}
\usepackage{siunitx}
\usepackage{amsfonts}
\usepackage{bm}
\usepackage{caption}
\usepackage{subcaption}
\usepackage{braket}
\usepackage{hyperref}
\usepackage{lineno}

\hypersetup{
    colorlinks=true,
    citecolor=blue,
    filecolor=blue,
    linkcolor=blue,
    urlcolor=blue
}

\usepackage[capitalize]{cleveref}

\begin{document}

\title{Sparse Reconstruction of Wavefronts using an Over-Complete Phase Dictionary}

\author{S. Howard}
\affiliation{Department of Physics, Clarendon Laboratory, University of Oxford, Oxford OX1 3PU, United Kingdom}
\affiliation{Ludwig-Maximilians-Universit{\"a}t M{\"u}nchen, Am Coulombwall 1, 85748 Garching, Germany}

\author{N. Weiße}
\affiliation{Ludwig-Maximilians-Universit{\"a}t M{\"u}nchen, Am Coulombwall 1, 85748 Garching, Germany}

\author{J. Schröder}
\affiliation{Ludwig-Maximilians-Universit{\"a}t M{\"u}nchen, Am Coulombwall 1, 85748 Garching, Germany}

\author{C. Barbero}
\affiliation{Grupo de Investigación en Aplicaciones del Láser y Fotónica (ALF), Universidad de Salamanca, 37008 Salamanca, Spain}

\author{B. Alonso}
\affiliation{Grupo de Investigación en Aplicaciones del Láser y Fotónica (ALF), Universidad de Salamanca, 37008 Salamanca, Spain}
\affiliation{Unidad de Excelencia en Luz y Materia Estructuradas (LUMES), Universidad de Salamanca, Spain}

\author{Í. Sola}
\affiliation{Grupo de Investigación en Aplicaciones del Láser y Fotónica (ALF), Universidad de Salamanca, 37008 Salamanca, Spain}
\affiliation{Unidad de Excelencia en Luz y Materia Estructuradas (LUMES), Universidad de Salamanca, Spain}

\author{P. Norreys}
\affiliation{Department of Physics, Clarendon Laboratory, University of Oxford, Oxford OX1 3PU, United Kingdom}

\author{A. Döpp}
\affiliation{Department of Physics, Clarendon Laboratory, University of Oxford, Oxford OX1 3PU, United Kingdom}
\affiliation{Ludwig-Maximilians-Universit{\"a}t M{\"u}nchen, Am Coulombwall 1, 85748 Garching, Germany}
\email{a.doepp@lmu.de}

\begin{abstract} 
Wavefront reconstruction is a critical component in various optical systems, including adaptive optics, interferometry, and phase contrast imaging. Traditional reconstruction methods often employ either the Cartesian (pixel) basis or the Zernike polynomial basis. While the Cartesian basis is adept at capturing high-frequency features, it is susceptible to overfitting and inefficiencies due to the high number of degrees of freedom. The Zernike basis efficiently represents common optical aberrations but struggles with complex or non-standard wavefronts such as optical vortices, Bessel beams, or wavefronts with sharp discontinuities. This paper introduces a novel approach to wavefront reconstruction using an over-complete phase dictionary combined with sparse representation techniques. By constructing a dictionary that includes a diverse set of basis functions — ranging from Zernike polynomials to specialized functions representing optical vortices and other complex modes — we enable a more flexible and efficient representation of complex wavefronts. Furthermore, a trainable affine transform is implemented to account for misalignment. Utilizing principles from compressed sensing and sparse coding, we enforce sparsity in the coefficient space to avoid overfitting and enhance robustness to noise. An implementation of the technique is publicly available\footnote{\href{https://github.com/sunnyhoward/overdictionary}{Over-Complete Dictionary}}.
\end{abstract}

\maketitle

\section{Introduction}

Wavefront sensing serves as a pivotal element in numerous domains ranging from computer vision and phase contrast imaging to optical imaging and astrophysics. The wavefront plays a crucial role in shaping the intensity distribution of light as it propagates. This fundamental relationship between phase and intensity forms the basis for various wavefront measurement techniques. As light travels through space or interacts with optical elements, the phase of the wavefront undergoes changes, which manifest as variations in the observed intensity patterns. By carefully analyzing these intensity variations, it becomes possible to infer the underlying phase information.

Wavefront measurement techniques leverage this principle in different ways. Some techniques, such as the Gerchberg-Saxton (GS) algorithm \cite{GS} or transport of intensity equation (TIE) methods\cite{zuo2020transport}, rely on knowledge of the intensity at a reference plane and compare it with intensities measured at other planes to reconstruct the phase. Others, like interferometric methods \cite{primot1995achromatic} or wavefront slope sensors\cite{SH-history}, manipulate the wavefront to create intensity patterns that encode the phase information.

The distinction between these techniques lies in their assumptions about the reference intensity and how they utilize the available intensity measurements. Some techniques require precise knowledge of the reference intensity, while others can work with relaxed assumptions or infer it from measurements in a different plane.

For any wavefront measurement technique, the wavefront is extracted from the raw signal using a reconstruction process. This process varies depending on the method - for example, it might involve propagation calculations in the Gerchberg-Saxton algorithm or the stitching of gradients in other methods. In this reconstruction step, an important consideration is the basis in which the wavefront is expressed.

The raw signal on the sensor is typically measured on a pixel grid arranged in Cartesian coordinates, often referred to as the pixel basis. Each pixel represents a mode in this basis, with its intensity value serving as the coefficient. Conceptually, it might seem straightforward to perform the reconstruction of the wavefront in this same basis, determining the phase value for each individual pixel. This approach, known as zonal reconstruction, is commonly used in various wavefront sensing methods, including Lateral Shearing Interferometry (LSI) and Shack-Hartmann (SH) sensors \cite{DAI2016264,Pathak_2014, southwell1980wave, chanteloup2005multiple,tian1995simple}.

The pixel basis maintains a direct correspondence between spatial positions in the raw signal and the reconstructed wavefront, making it well-suited for capturing high-frequency features such as edges. This has made it prominent in applications like computer vision and phase contrast imaging. However, the pixel basis has a significant limitation: its proneness to overfitting. Given the high degrees of freedom in the Cartesian basis - typically one value per pixel - there is an increased risk of modeling noise instead of the actual signal, especially in the presence of limited data or high noise levels.

These limitations of the pixel basis motivate the exploration of alternative representations for wavefront reconstruction that can leverage prior knowledge about the optical system to mitigate overfitting and provide more robust reconstructions.

\begin{figure}
    \centering
    \includegraphics[width=\linewidth]{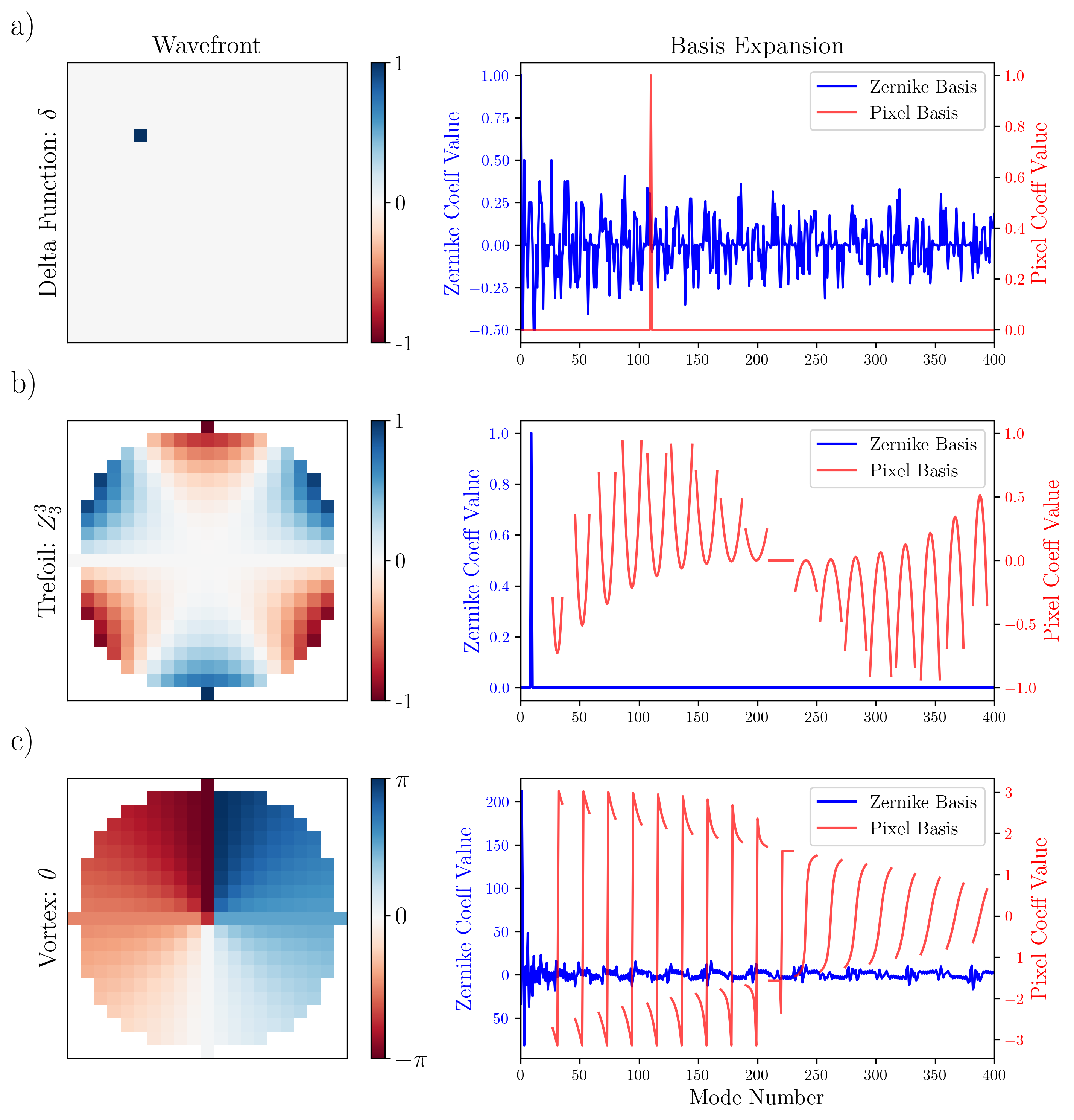}
    \caption{The expansion of a number of wavefronts in terms of the Pixel and Zernike bases. In order to describe wavefront function, $\Psi$, the coefficient for mode $n$ in basis $\Phi$ was calculated by $c_n = \braket{\Psi|\Phi_{n}}$. a) To express a single pixel requires many Zernike modes. b) A single Zernike mode requires many modes in the Pixel basis. c) There exist wavefronts, such as the optical vortex, which are not efficiently represented in either the Zernike or Pixel basis. The Zernike modes are ordered according to their Noll index \cite{noll1976zernike}, and the colorbar applies for all wavefronts.}
    \label{fig:enter-label}
\end{figure}

With prior knowledge about the system, one can choose a more suitable basis that avoids the problems of overfitting noise; for example, the often used Zernike basis\cite{liang1994objective,He:21, seifert2005wavefront, howard2023hyperspectral, harbers1996analysis, dai2016modal}. The Zernike basis can be intuitively understood as a polar coordinate adaptation of polynomial regression, optimized for representing wavefront aberrations in circular optical systems. By design it is directly related to common optical surface aberrations like coma and astigmatism, making it highly suitable to describing wavefronts that were generated with optical components. Defined on a unit disk, the Zernike polynomials provide an efficient way to handle rotationally symmetric systems, common in optical applications. The Zernike modes are also somewhat hierarchical and by truncating the coefficients one can reliably calculate standard aberrations. For example, a laser wavefront can typically be described by a number of Zernike modes on the order of 10, which represents many fewer coefficients than required in the pixel basis. However, the efficiency of the Zernike basis is diminished when confronted with more complex or non-standard aberrations, for example those with sharp edges; \cref{fig:enter-label}a-b) displays the benefit of employing the Cartesian or Zernike basis in certain scenarios.

There are other wavefronts which are not efficiently captured by either the Zernike or Pixel basis, such as optical vortices \cite{vortex}, which possess an azimuthally varying phase that remains radially constant. The phase profile of an optical vortex can be written as $\phi(\theta) = \ell\theta$, where $\ell$ is the topological charge and $\theta$ is the azimuthal angle. Despite its very simple form, this phase profile cannot be efficiently captured by either Cartesian or Zernike polynomials, a fact demonstrated in \cref{fig:enter-label}c). In addition to optical vortices, myriad other complex wavefront morphologies, such as Bessel beams\cite{mcgloin2005bessel}, Laguerre-Gaussian beams, and Airy beams\cite{efremidis2019airy}, each carrying unique phase and amplitude profiles, are often encountered in optics. Each of these can be captured via concise mathematical descriptions, but require complex expressions in other bases. Therefore, it is apparent that a more versatile and encompassing framework is required for wavefront reconstruction.

Over-complete dictionaries offer a promising solution to this challenge. In contrast to traditional basis sets which are typically complete and orthogonal, over-complete dictionaries contain a surplus of basis functions. While traditional basis sets allow each signal to be represented in one unique way, over-complete dictionaries offer multiple representations for the same signal, providing additional flexibility. Adopting methods from compressed sensing and sparse signal representation, the L1 norm can be employed to enforce sparsity during the reconstruction process, i.e., favoring solutions with mostly zero-valued elements except for a few significant components. This approach enables the finding of the most efficient representation of the waveform.

Capitalizing on this theory, we propose the construction of an over-complete phase dictionary, a versatile set of basis functions that extend beyond the capabilities of Cartesian and Zernike polynomials. The proposed dictionary, rich in the variety of basis functions it offers, aims to efficiently handle a wide array of wavefront morphologies.

This paper explores the design of the over-complete phase dictionary, delves into the sparse reconstruction methodology using the dictionary, and its potential implications. This way, we provide a more accurate, interpretable, and computationally efficient representation of wavefronts, pushing the boundaries of traditional wavefront reconstruction methodologies.

\section{Theory}

Traditional methodologies for wavefront reconstruction have primarily relied on two types of basis sets, known as zonal and modal reconstruction, respectively: the Cartesian and Zernike basis \cite{tyson2022principles}. While each has its merits, they both possess inherent limitations that render them inadequate for capturing complex wavefronts. The constraints observed in these approaches call for the formulation of a comprehensive, versatile basis set that efficiently handles a broad array of wavefront morphologies. 

The following considers the reconstruction of the wavefront from measurements of its gradient, $\vec{s}$, which is the relevant reconstruction process for techniques such as the Shack-Hartmann sensor and lateral shearing interferometry. Using a finite difference derivative matrix, $\mathbf{D}$, the forward process can be described using the vectorized form of the wavefront, $\vec{w}$, as, $\vec{s} = \mathbf{D}\vec{w}$. The wavefront can be expressed in any complete basis via, $\vec{w} = \mathbf{B}\vec{c} = \sum_{i} \vec{(B_{i})}c_i$, where $\vec{c}$ and $\mathbf{B}$ are the modal coefficients and basis functions respectively. Combining these equations, and defining a new matrix as the derivatives of a set of basis functions, $\mathbf{D_{B}}$, one arrives at, 
\begin{equation}
    \vec{s} = \mathbf{D_{B}}\vec{c}.
\end{equation}
The reconstruction task is to find the coefficients, $\vec{c}$, given $\vec{s}$ and $\mathbf{D_{B}}$. The key differences in existing approaches center on the choice of basis functions, $\mathbf{B}$, as is discussed in the following section.

\subsection{Zonal and Modal Reconstruction}

Zonal reconstruction represents the wavefront using a piecewise function, typically in the Cartesian basis. This approach excels at resolving high-frequency spatial features but suffers from issues like overfitting and sensitivity to noise.

\begin{equation}
W(x, y) = \sum_{i=0}^{n} \sum_{j=0}^{m} c_{ij} \delta(x-i)\delta(y-j)
\end{equation}

Here, \( W(x, y) \) is the wavefront, and \( c_{ij} \) are the coefficients.

Modal reconstruction, on the other hand, employs global functions, such as Zernike polynomials, that span the entire aperture. This approach is particularly effective for describing rotationally symmetric systems and is less susceptible to overfitting. A wavefront is expanded in the Zernike polynomials, \( Z_{nm} \), according to,

\begin{equation}
W(R, \theta) = \sum_{n=0}^{\infty} \sum_{m=-n}^{n} c_{n}^{m} Z_{nm}(R, \theta),
\end{equation}

where $R$ and $\theta$ are the radial coordinate and polar angle respectively, and \( c_{n}^m \) are the corresponding coefficients.

\subsection{Limitations of Zonal and Modal Bases}

Each basis set has its limitations. The Cartesian basis is often inefficient for low-frequency components due to its high dimensionality. Conversely, Zernike polynomials struggle to capture high-frequency details or non-standard aberrations. As a result, researchers are often forced to compromise between robustness and representational power. In the following two simple examples are discussed, in which the wavefront can be easily parameterized under a specific basis, but is inefficiently encoded in the common zonal and modal approaches.

\subsubsection{Optical Vortex}

Consider an optical vortex, with a wavefront described by $\ell\theta$. Trivially, in the Cartesian (pixel) basis, one requires a number of elements equal to the number of pixels to describe this wavefront. Less obvious is that the Zernike basis also requires more coefficients. As the Zernike polynomials form a complete basis, one can expand any function in this basis, 
\begin{equation}
    f(\theta,R) = \sum_{m,n} c_n^m Z_n^m
\end{equation}

To find the expansion coefficients, one simply calculates the overlap with the given Zernike and the function over the domain which Zernikes are orthogonal, as can be seen below;

\begin{equation}
        \int_{-\pi}^{\pi} f(\theta, R) Z_n^m = \int_{-\pi}^{\pi} \left(\sum_{m',n'} c_{n'}^{m'} Z_{n'}^{m'}\right) Z_n^m = c_n^m
\end{equation}

Setting the function equal to the vortex, \( f(\theta, R) = \ell \theta \), one may attempt to expand it in terms of Zernike polynomials:

\begin{equation}
\ell \theta = \sum_{n=0}^{\infty} \sum_{m=-n}^{n} c_n^m Z_n^m(R, \theta)
\end{equation}

Detailed analysis (see appendix) shows that one needs the complete set of coefficients \( c_n^m \), each proportional to \( 1/m \) in magnitude. Thus, the representation of this seemingly simple function is very inefficient. In practice, this problem is amplified by the fact that higher order Zernike modes are numerically unstable. This can only to some extent be mitigated using recursive definitions\cite{andersen2018efficient}, making it impractical to describe wavefronts with more than Zernike modes. 

\subsubsection{Misaligned Zernike Modes}\label{sec:misaligned}

Consider a wavefront that can be represented by a single Zernike coefficient, e.g. the oblique trefoil mode, $Z^{3}_3(x,y)$, which is described in polar and cartesian coordinates as, 
\begin{equation}
    Z^{3}_3 = R^3 \cos(3\theta) = x^{3} - 3xy^2
\end{equation}

Now consider a small shift, $x \rightarrow x + \Delta$, so the Zernike mode is offset from the center.
\begin{widetext}
\begin{align}
    Z^{3}_3 (x+\Delta,y) &= (x + \Delta)^3 - 3(x + \Delta)y^2\\
    &= x^3 - 3xy^2  + x^2(-3\Delta) + x(3\Delta ^2) - \Delta^3 - 3\Delta y^2\\
    &= Z^{3}_3 (x,y)  + \underbrace{(x^2 - y^2)(-3\Delta)}_\text{n=2} + \underbrace{x(3\Delta ^2)}_\text{n=1} - \underbrace{\Delta^3}_\text{n=0} 
\end{align}
\end{widetext}
This demonstrates that a simple coordinate shift of a Zernike coefficient in row $N$ of the pyramid will introduce additional terms from rows $[0,...,N-1]$; a basis that was sparse will now be very dense. One also notes for a row, $n$, the shift factor is proportional to $\Delta^{N-n}$. As $\Delta < 1$ (the Zernike basis is on a unit circle), the higher order modes are added with greater weighting. To generalize this observation, one employs a Taylor expansion of the shifted Zernike mode:
\begin{align} 
    Z^{M}_N (x+\Delta, y) &= Z^{M}_N (x,y) + \Delta\frac{\partial Z^{M}_N}{\partial x} +\frac{\Delta^2}{2!}\frac{\partial^2 Z^{M}_N}{\partial x^2} + \cdots
\end{align} 
The derivatives of the Zernike polynomials can be expressed as linear combinations of Zernike polynomials of lower order. Therefore, the first derivative is
\begin{equation}
    \frac{\partial Z_N^M}{\partial x} = \sum_{n = N - 1}^0 \sum_{m} a_{n m} Z_n^m(x, y)
\end{equation}
and similarly, higher-order derivatives involve even lower-order polynomials. Substituting back into the Taylor expansion, one obtains:
\begin{align}
    Z_N^M(x + \Delta, y) &= Z_N^M(x, y) + \Delta \sum_{n = N - 1}^0 \sum_{m} a_{n m} Z_n^m(x, y) \nonumber \\
    &\quad + \frac{\Delta^2}{2} \sum_{n = N - 2}^0 \sum_{m} b_{n m} Z_n^m(x, y) + \cdots
\end{align}
This series demonstrates that a small shift \( \Delta \) introduces contributions from lower-order Zernike modes down to \( n = 0 \). The coefficients \( a_{n m} \) and \( b_{n m} \) depend on the specific mode and its derivatives. Even though \( \Delta \) is small, the impact on the representation is significant because the higher-order terms decay slowly due to the factorial in the denominator being offset by the increasing number of contributing lower-order modes.

Thus, a wavefront that was sparsely represented by a single Zernike coefficient when centered becomes densely represented upon shifting. This density arises from the need to account for the introduced lower-order modes to accurately describe the shifted wavefront. The practical implication is that even minor misalignments in optical systems, or in the detection process, can lead to significant inefficiencies in wavefront representation and interpretation when using Zernike polynomials.

This effect underscores the limitations of the Zernike basis in situations where the wavefront is not perfectly centered or aligned, which is often possible in the case of an imperfectly-circular beam. It highlights the necessity for a more adaptable representation that maintains sparsity despite shifts or misalignments.

\subsection{Over-Complete Dictionaries}

Over-complete dictionaries\cite{chen2001atomic, rubinstein2010dictionaries} provide a solution to the problems outlined above. These dictionaries contain more basis functions than the dimensions in the space they represent, offering flexibility to adapt to a variety of signals while maintaining robustness.

\begin{equation}
W(x,y) = \sum_{i=1}^{N} c_i \phi_i(x,y)
\end{equation}

Here, \( \phi_i \) are the basis functions from the over-complete dictionary, and \( c_i \) are the coefficients. The overcompleteness of our dictionary presents a fundamental challenge: the existence of multiple valid solutions for a given wavefront. This non-uniqueness might seem to undermine the very purpose of our approach. However, it is precisely this redundancy that allows one to overcome the limitations of single-basis representations that have been previously discussed.

The pursuit of sparsity in this context is not merely a mathematical convenience, but a reflection of the underlying physics of wavefronts. Throughout this paper, it has demonstrated that different physical phenomena in optics - be it Zernike-type aberrations, optical vortices, or misaligned beams - naturally lend themselves to sparse representations in appropriate bases. This sparsity is a consequence of the wave equation and the boundary conditions typical in optical systems.

The limitations of individual bases in efficiently representing diverse wavefront phenomena stem from the fundamental nature of optical wavefronts. These are in most cases solutions to the paraxial wave equation:
\begin{equation}
\frac{\partial^2 \psi}{\partial x^2} + \frac{\partial^2 \psi}{\partial y^2} + 2ik\frac{\partial \psi}{\partial z} = 0
\end{equation}
where $\psi(x,y,z)$ is the complex amplitude of the field, $k$ is the wavenumber, and $z$ is the propagation direction. The solutions to this equation encompass a wide range of phenomena, including Gaussian beams, Hermite-Gaussian modes, Laguerre-Gaussian modes (which include optical vortices), and Bessel beams, among others.

The pursuit of sparsity in our over-complete dictionary is fundamentally a search for the most appropriate set of solutions to the paraxial wave equation that describe the observed wavefront. This approach can be formalized mathematically as an optimization problem:
\begin{equation}
\min_{\mathbf{c}} |\mathbf{c}|_0 \quad \text{subject to} \quad W = \Phi\mathbf{c}
\end{equation}
Here, $|\mathbf{c}|_0$ denotes the L0 norm, which quantifies the number of non-zero elements in the coefficient vector $\mathbf{c}$, and $\Phi$ is the dictionary of modes. This formulation seeks to identify the minimal set of fundamental modes that accurately represent the observed wavefront.

However, L0 minimization is NP-hard, leading us to the more tractable L1 norm minimization, as proposed by Cand{\`e}s et al. \cite{candes2006robust}:

\begin{equation}
\min_{\mathbf{c}} \|\mathbf{c}\|_1 \quad \text{subject to} \quad W = \Phi\mathbf{c}
\label{l1_reg}
\end{equation}

This relaxation, while computationally necessary, still promotes sparsity and often yields solutions very close to the true L0-sparse solution. Cand{\`e}s and colleagues demonstrated that under certain conditions, L1 minimization can exactly recover the sparsest solution, providing a theoretical foundation for compressed sensing and sparse signal reconstruction \cite{candes2006robust,candes2006near}.

Crucially, the paraxial wave equation is defined with respect to an optical axis, which in real optical systems may not perfectly align with the measurement apparatus. As discussed in Sec.\ref{sec:misaligned}, this misalignment can lead to apparent complexity in the wavefront when viewed from the perspective of the measurement system. By including the center coordinates, $(\Delta_x, \Delta_y)$, as variables in our optimization, one may account for this potential misalignment:
\begin{equation}
\min_{\mathbf{c}, \Delta_x, \Delta_y} |\mathbf{c}|_0 \quad \text{subject to} \quad W = \Phi(x-\Delta_x, y-\Delta_y)\mathbf{c}
\end{equation}
This formulation allows us to simultaneously determine the most appropriate set of paraxial modes and the true optical axis of the system. It recognizes that apparent complexity in the wavefront may arise from a simple misalignment rather than intrinsic high-order aberrations or complex phase structures.

When we introduce the idea of fitting the center during reconstruction, we are implicitly expanding our dictionary to include shifted versions of each basis function. Mathematically, we can express this as:
\begin{equation}
\Phi_{expanded} = \{ \phi_i(x-\Delta_x, y-\Delta_y) : \phi_i \in \Phi, (\Delta_x, \Delta_y) \in \mathbb{R}^2 \}
\end{equation}
This expanded dictionary is infinite-dimensional, as it includes all possible shifts of each basis function. However, it's important to note that this expansion doesn't fundamentally change the nature of our problem - it remains an over-complete dictionary, just with an even higher degree of redundancy.

The key insight is that by allowing for these shifts, we're enabling our reconstruction to find even sparser representations. A misaligned Zernike mode, which might require many coefficients in the original dictionary, can now be represented by a single coefficient in the expanded dictionary.

Incorporating this into our optimization problem using the L1 norm, we get:
\begin{equation}
\min_{\mathbf{c}, \Delta_x, \Delta_y} \|\mathbf{c}\|_1 \quad \text{subject to} \quad W = \Phi(x-\Delta_x, y-\Delta_y) \mathbf{c}
\end{equation}
This formulation connects seamlessly with the theory of over-complete dictionaries and sparse representations. We're still seeking the sparsest representation, but now in an expanded dictionary that can capture a wider range of physical phenomena efficiently.

The concept of dictionary learning in sparse coding literature provides a theoretical framework for understanding this approach. Just as dictionary learning algorithms adapt the basis functions to better represent a class of signals, our method adapts the positioning of the basis functions to better represent the specific wavefront at hand.

This adaptive approach not only allows for more accurate reconstruction of misaligned wavefronts but also provides valuable information about the optical system itself. The optimal $(\Delta_x, \Delta_y)$ values can indicate misalignments in the optical setup or in the detection process, offering diagnostic capabilities beyond mere wavefront reconstruction.

We refer to our technique by the acronym: PROD (Phase Reconstruction using an Over-complete Dictionary). In the following sections, we will present numerical experiments demonstrating how this expanded dictionary approach, combined with L1-based sparse reconstruction, outperforms static reconstructions across a wide range of wavefront types. We'll explore its robustness to various aberrations and misalignments, showcasing how the pursuit of sparsity in an appropriately designed over-complete dictionary can lead to both more accurate and more interpretable wavefront reconstructions.

\begin{figure*}[t]
    \centering
    \includegraphics[width=.98\linewidth]{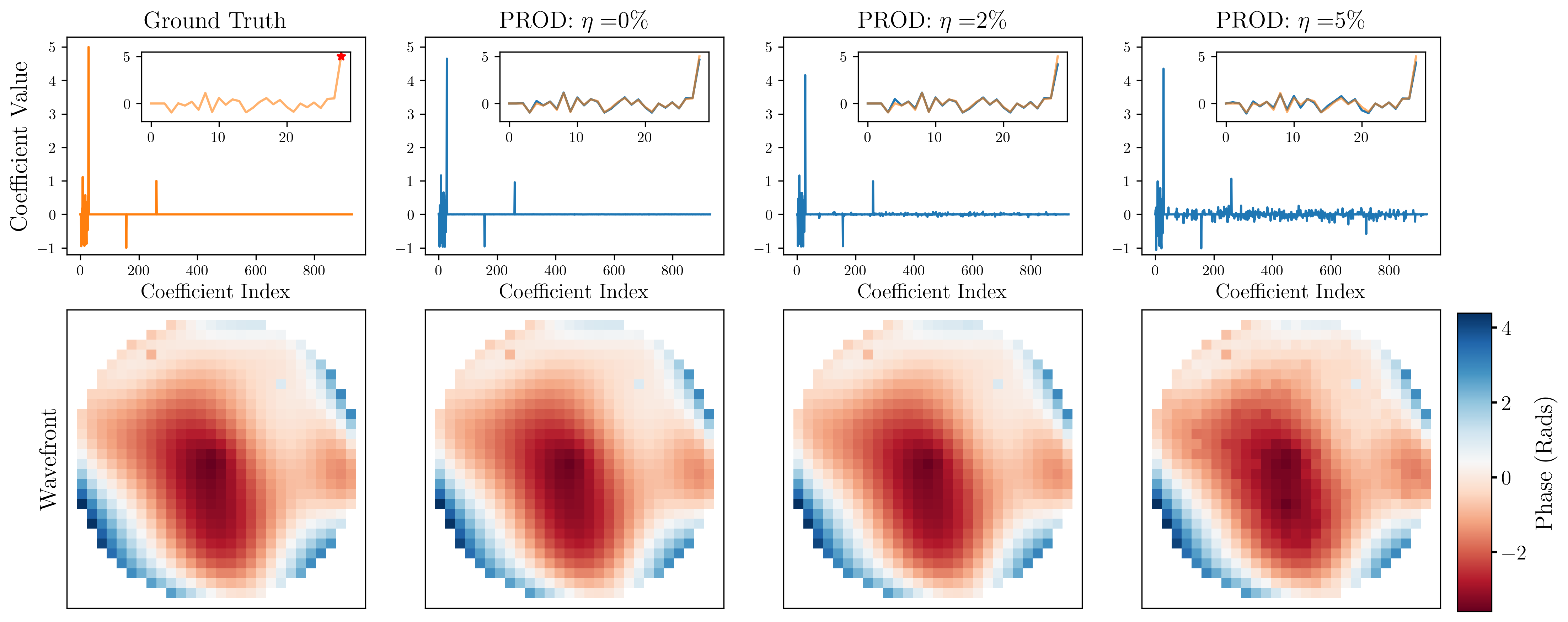}
    \caption{The wavefront was synthesized by the random initialization of 28 Zernike modes, and the Axicon mode. The top row displays the modal coefficient values, with each subplot showing a zoom of the first 29 coefficients - the Zernikes and Axicon (marked with a red star). The bottom row shows the predicted wavefront, with all plotted on the same color scale. One sees that even in the presence of significant noise (parameterized by a percentage, $\eta$, of the maximum value of the derivative), PROD is able to accurately extract the true coefficient values. }
    \label{fig:axicon}
\end{figure*}

\section{Numerical Simulations}
\subsection{Construction of the Over-Complete Dictionary}

Given a set of wavefront derivatives, with spatial dimensions $(N_x, N_y)$, the over-complete dictionary is created by the synthesis of a number of modes from different bases. Added first is the pixel basis, with the introduction of $N_x\times N_y$ modes, allowing the technique to represent any discontinuities. Then, to provide an efficient way to represent the common low frequency features, a set number of Zernike modes are added, where the derivatives are found analytically \cite{andersen2018efficient}. Finally, with prior knowledge of the optical system being measured, one can simply add `special' modes that are expected to be present in the wavefront, such as the optical vortex ($\phi = \theta)$. The topological charge of the vortex will then be the modal coefficient that is found by the PROD. As has been described, the pixel basis is already sufficient to describe any wavefront. However, in doing so, it will also capture noise, and it does not provide useful physically interpretable modal coefficients; for example the order of a vortex.

Once the over-complete dictionary has been formed by the concatenation of the modal derivatives, a spatial transformer network is also initialized \cite{jaderberg2015spatial}. This is a module which performs a translation of the Zernike modes and an affine transform of the special modes, with trainable parameters for the rotation angle, $\theta$, and translations, ($\Delta_x, \Delta_y$), enabling the most efficient representation of the wavefront. The Zernike modes and the special modes are each given their own set of affine parameters, to account for the potentially different sources of these distortions. The trainable modal coefficients and affine parameters are then optimized by minimizing the error with the measured wavefront derivatives whilst simultaneously minimizing the L1 norm of the coefficients, according to \cref{optimization}.

In practice, at initialization we create multiple sets of random coefficients and random (or grid) affine parameters, and optimize them in parallel. By then choosing the one that achieved the minimum cost, the robustness of the technique is increased, without a dramatic increase in computational time due to the properties of GPUs. Due to the affine transformation, the spatial domain which the over-complete dictionary is defined on is larger than the original derivatives, and depends on the upper bounds of the translation and rotation parameters; during optimization, the predicted wavefront derivatives are simply cropped to the same size as the measured ones for the cost calculation.

\begin{figure*}[pt]
    \centering
    \includegraphics[width=.95\linewidth]{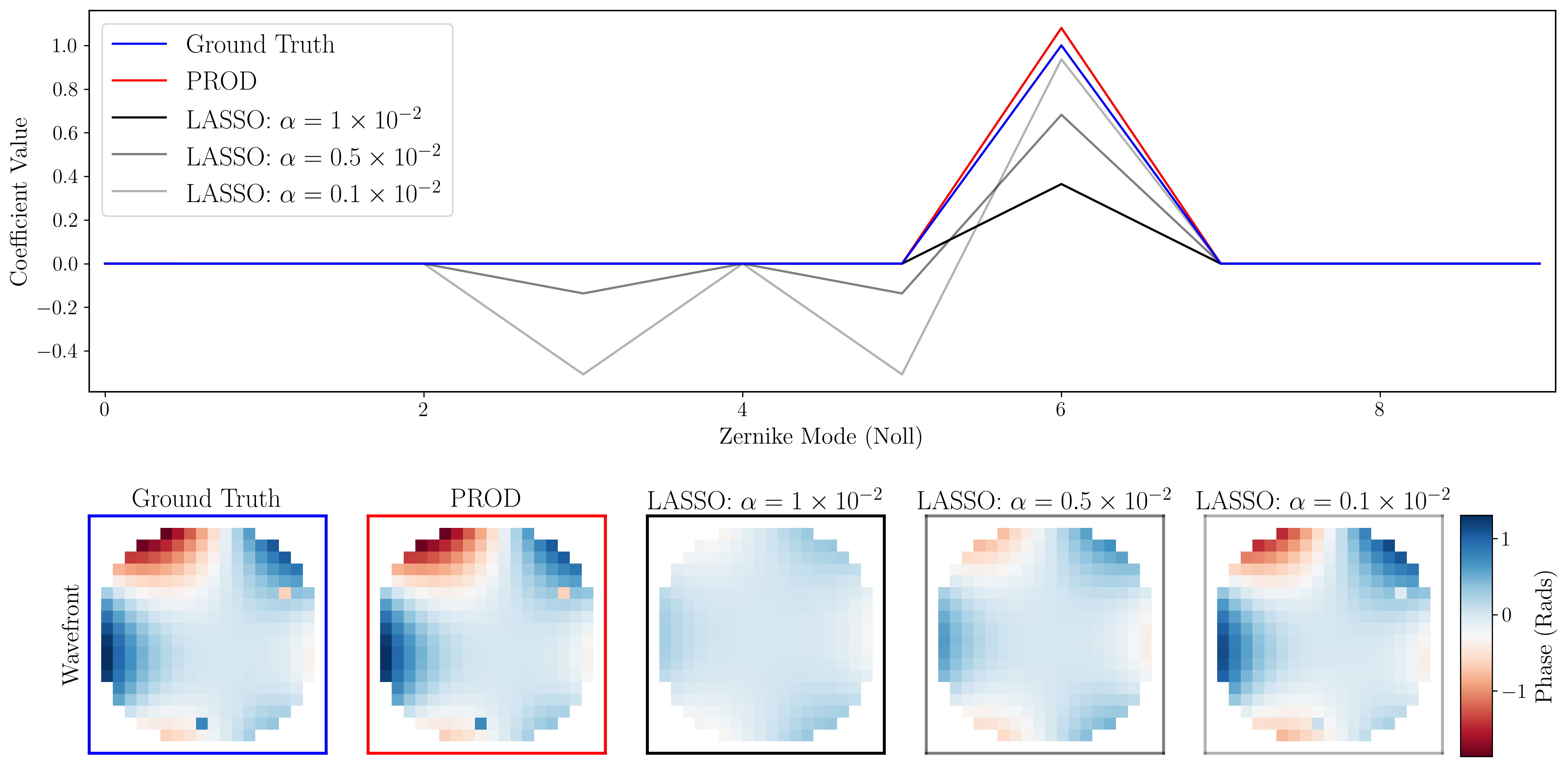}
    \caption{Demonstrating the utility of PROD's affine transform module to efficiently represent de-centered modes, in this case the $Z_{3}^{3}$ mode. Here PROD learns the de-center parameter, and identifies the single Zernike mode. Also displayed for comparison is the LASSO technique, which is given the same over-complete dictionary, but possesses no ability to transform the modes. All wavefronts are plotted on the same color scale. If its L1 penalization parameter, $\alpha$, is large, it cannot represent the wavefront accurately, whereas if $\alpha$ is small, it must utilize lower order Zernike modes to help represent the wavefront.}
    \label{fig:LassoComp}
\end{figure*}

\subsection{Optimization Framework}\label{optimization}

While L1 minimization problems are traditionally approached using convex optimization algorithms like Iterative Shrinkage-Thresholding Algorithms (ISTA)\cite{beck2009fast}, Interior Point Methods (IPM) \cite{kim2007interior} or the Alternating Direction Method of Multipliers (ADMM) \cite{boyd2011distributed}, we opt for the Adaptive Moment Estimation (ADAM) optimizer \cite{kingma2014adam} implemented in PyTorch.

The complexity of our optimization problem, particularly with the introduction of center fitting, makes it potentially non-convex. ADAM's ability to handle both convex and non-convex problems effectively makes it a suitable choice. Moreover, ADAM's efficient GPU-accelerated implementation in frameworks like PyTorch can in practice outperform traditional convex optimization algorithms without similar optimization.

We formulate our sparse optimization problem as:
\begin{equation}
\min_{\mathbf{c}, \Delta_x, \Delta_y} \|\mathbf{W} - \Phi(x-\Delta_x, y-\Delta_y)\mathbf{c}\|_2^2 + \lambda\|\mathbf{c}\|_1 \label{cost}
\end{equation}
where $\mathbf{W}$ is the measured wavefront (or more precisely its slopes), $\Phi$ is our over-complete dictionary, $\mathbf{c}$ are the coefficients we're optimizing, $\delta_x$ and $\delta_y$ are the center coordinates, and $\lambda$ is a regularization parameter. This approach combines the flexibility to handle potential non-convexity, the speed of GPU acceleration, and the sparsity-inducing properties required for our over-complete wavefront reconstruction problem. The following experiments were performed on a single NVIDIA GeForce RTX 3090 GPU.

\subsection{Results}

\begin{figure*}[t]
    \centering
    \includegraphics[width=0.95\linewidth]{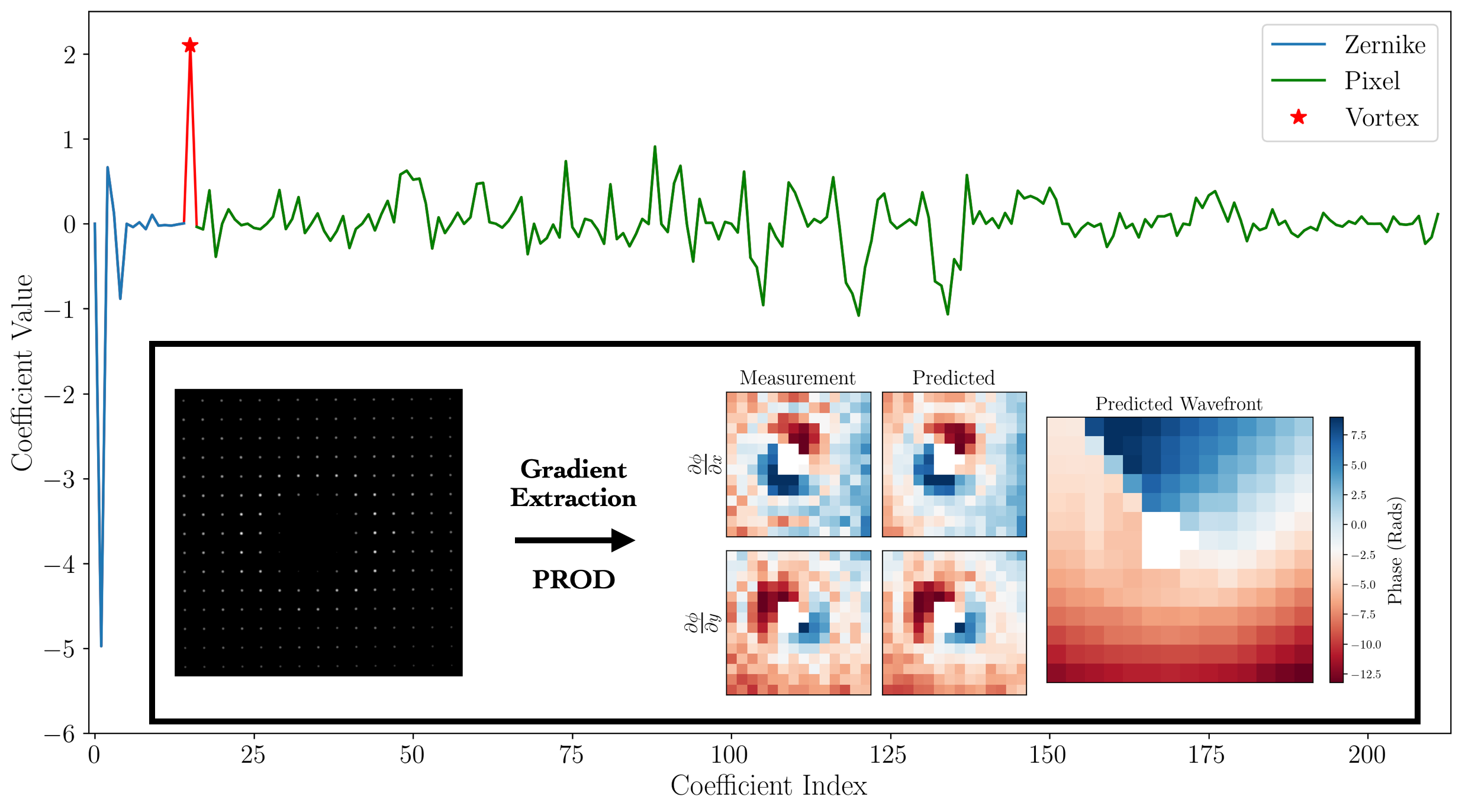}
    \caption{The use of PROD to characterize an experimental optical vortex. Once the gradients are extracted from the Shack-Hartmann pattern, the technique is used to find the modal coefficients for the vortex, and lower order Zernike modes. The coefficient for the vortex (the topological charge) was 2.09, in good agreement with the theoretical value for this setup of 2.}
    \label{fig:physical_measurement}
\end{figure*}

\subsubsection{Axicon Phase}

Firstly, PROD is utilized to perform phase stitching of a wavefront not efficiently expressed in the Zernike basis: a simple linear radial phase, $\phi(r) \propto r$. Such a wavefront can be generated by an Axicon \cite{axicon}, a type of lens formed from a conical piece of glass. Due to its geometry, there is a linear dependence in the radial position and the amount of glass that the pulse has traveled through, resulting in the radial wavefront. When used to focus a Gaussian beam, it creates a Bessel-like beam which does not experience diffraction over a region of interest \cite{garces2002simultaneous}, making it of interest in a wide range of applications. 

Here, we create a wavefront by randomly sampling Zernike coefficients from a Normal distribution $\sim \mathcal{N}(0,0.5)$, up to the 28th mode, before adding the Axicon phase. To test the ability of the PROD technique to handle noise, 3 sampling scenarios are simulated. Normally distributed noise is added to the gradients, with a standard deviation equal to a percentage, $\eta \in [0\%, 2\%, 5\%]$, of the gradients maximum value. 

The learning rate of the ADAM optimizer was set to $3\times 10^{-3}$, and the regularization parameter was set to $\lambda = 7\times 10^{-4}$.  The results are shown in \cref{fig:axicon}.

It is evident that as the noise level was increased, PROD was able to use the pixel basis in order to fit the noise, meaning that the predictions of the Zernike and Axicon mode remained fairly constant throughout the noise levels, proving the robustness of the technique. 

\subsubsection{Shifted Oblique Trefoil}

This experiment demonstrates the benefits of the affine transform aspect of the approach. The phase map is implemented as an off-centered $Z_{3}^{3}$ Zernike mode (oblique trefoil), with 2 pixels of shot noise. Considering the Zernike modes are defined from [-1,1], the new center position is chosen as [0.2, 0.2]. Normally distributed random noise ($\eta = 2\%$) was added to the sampled derivatives, before the proposed approach was used to recover the wavefront from the derivatives. For comparison, we use the scikit-learn\cite{scikit-learn} implementation of LASSO regression\cite{lasso}. This is a commonly implemented technique to solve \cref{l1_reg}, and is described in the appendix. The technique is given the same over-complete dictionary, but possesses no ability to perform the transformation of modes. LASSO is parameterized by a L1 penalization parameter, $\alpha$, which controls the sparsity of the solution. This was adjusted and the results are displayed in \cref{fig:LassoComp}. 

As alluded to in \cref{sec:misaligned}, a shifted Zernike mode introduces other lower order Zernike terms into the expansion. This is seen for the LASSO technique; as $\alpha$ is decreased, more low order Zernike modes are used to help represent the function. As the proposed approach uses a trainable affine transform, it finds the most efficient representation using just the single Zernike mode.

\section{Physical Experiments}

An experiment was performed to measured an optical vortex. This  was performed using a Spectra-Physics Spitfire ACE laser system to generate ultrashort pulses with central wavelength, $\lambda_0 = 798$ nm, and a transform limited duration of 64 fs. To generate the vortex, a setup was used that has been thoroughly discussed in previous work \cite{swaveplate, lopez2022bulk}. Firstly, a quarter waveplate (QWP) was used to circularly polarize the incoming pulse, before it passed through a $(l=2)$ structured waveplate (s-waveplate), creating two circularly polarized vortices with opposing handedness. A further QWP was used to transform them into two linearly polarized vortices before a linear polarizer was used to select one of them. The optical components used here were achromatic, so the signal was integrated over time when captured. The wavefront measurement was then performed on a home-built Shack-Hartmann sensor (focal length, $f = 14.2$ mm, and pitch, $\Lambda$ = $300$ µm) \cite{nils}. The microlens focii for the vortex are seen in \cref{fig:physical_measurement}. 

A reference microlens focii pattern was first captured, before the s-waveplate was introduced to the optical setup. The centroids of each microlens focii were extracted, before the wavefront gradients were found by calculating the centroid difference between the vortex and the reference measurements. The PROD technique was then used to extract the modal coefficients and the wavefront. The results are displayed in \cref{fig:physical_measurement}. We see that the PROD technique was able to correctly identify the 2nd order vortex, with a predicted modal coefficient of 2.09. This discrepancy of 5\% relative to the theoretical value can be attributed to misalignment in the distance between the sensor and the microlens array, and imperfection in the vortex generation.

\section{Conclusions and Outlook}

Leveraging an over-complete dictionary and trainable affine transforms, PROD has been introduced as a robust technique to reconstruct a wavefront from measurements of its gradients. Where most existing techniques typically reconstruct in the pixel or the Zernike basis, there exist limitations with both however; the former overfits to noise and doesn't give physically interpretable coefficients, and there are some specialized wavefronts that cannot be efficiently expressed in the Zernike basis, such as the optical vortex. Exploiting one's prior knowledge about a measurement, PROD is able to extract the physically-meaningful modal coefficients for the modes that are within its over-complete dictionary. Furthermore, as its dictionary also incorporates the pixel basis, it can express noise without influencing the other modal coefficients, such as those for Zernikes and special modes. 

In its presented form, a PROD reconstruction of a $20\times 20$ wavefront takes on the order of seconds, meaning it is principally for post-analysis. However, if one assumes that all modes are centered and removes the affine transform module of this technique, LASSO can be used to solve the L1 problem in a far shorter time (ms), allowing a real-time reconstruction for applications in adaptive optics. While this may be suitable in some cases, such as when one is certain that the system is centered, in the most general case this is not a suitable assumption, for example with a decentered vortex, and can lead to an under-sparse representation, and resultingly inaccurate modal coefficients. 

This application of overcomplete dictionaries and the principles of sparsity can also be applied to other problems in laser physics. For example, the reconstruction of the spectral phase from measurements can also be performed in different bases, and sometimes involves switching between bases\cite{alonso2020compact, lopez2024few}. Furthermore, with sufficient computational resources, there is no reason why this technique cannot be extended to recent work in the spatiospectral regime, where the wavefront is considered in three dimensions\cite{nils}.\\

\section*{Acknowledgements}
We would like to acknowledge useful discussions with the groups of Professor Peter Norreys and Dr. Andreas Döpp.

This work was supported by the Independent Junior Research Group "Characterization and control of high-intensity laser pulses for particle acceleration", DFG Project No.~453619281. We would also like to acknowledge UKRI-STFC grant ST/V001655/1, and the following funding sources: European Regional Development Fund and Consejería de Educación, Junta de Castilla y León (SA136P20); Ministerio de Ciencia e Innovación (PID2020-119818GB-I00, PID2023-149836NB-I00).

\bibliographystyle{apsrev4-2}
\bibliography{refs}

%apsrev4-2.bst 2019-01-14 (MD) hand-edited version of apsrev4-1.bst
%Control: key (0)
%Control: author (72) initials jnrlst
%Control: editor formatted (1) identically to author
%Control: production of article title (-1) disabled
%Control: page (0) single
%Control: year (1) truncated
%Control: production of eprint (0) enabled
\begin{thebibliography}{40}%
\makeatletter
\providecommand \@ifxundefined [1]{%
 \@ifx{#1\undefined}
}%
\providecommand \@ifnum [1]{%
 \ifnum #1\expandafter \@firstoftwo
 \else \expandafter \@secondoftwo
 \fi
}%
\providecommand \@ifx [1]{%
 \ifx #1\expandafter \@firstoftwo
 \else \expandafter \@secondoftwo
 \fi
}%
\providecommand \natexlab [1]{#1}%
\providecommand \enquote  [1]{``#1''}%
\providecommand \bibnamefont  [1]{#1}%
\providecommand \bibfnamefont [1]{#1}%
\providecommand \citenamefont [1]{#1}%
\providecommand \href@noop [0]{\@secondoftwo}%
\providecommand \href [0]{\begingroup \@sanitize@url \@href}%
\providecommand \@href[1]{\@@startlink{#1}\@@href}%
\providecommand \@@href[1]{\endgroup#1\@@endlink}%
\providecommand \@sanitize@url [0]{\catcode `\\12\catcode `\$12\catcode
  `\&12\catcode `\#12\catcode `\^12\catcode `\_12\catcode `\%12\relax}%
\providecommand \@@startlink[1]{}%
\providecommand \@@endlink[0]{}%
\providecommand \url  [0]{\begingroup\@sanitize@url \@url }%
\providecommand \@url [1]{\endgroup\@href {#1}{\urlprefix }}%
\providecommand \urlprefix  [0]{URL }%
\providecommand \Eprint [0]{\href }%
\providecommand \doibase [0]{https://doi.org/}%
\providecommand \selectlanguage [0]{\@gobble}%
\providecommand \bibinfo  [0]{\@secondoftwo}%
\providecommand \bibfield  [0]{\@secondoftwo}%
\providecommand \translation [1]{[#1]}%
\providecommand \BibitemOpen [0]{}%
\providecommand \bibitemStop [0]{}%
\providecommand \bibitemNoStop [0]{.\EOS\space}%
\providecommand \EOS [0]{\spacefactor3000\relax}%
\providecommand \BibitemShut  [1]{\csname bibitem#1\endcsname}%
\let\auto@bib@innerbib\@empty
%</preamble>
\bibitem [{\citenamefont {Gerchberg}(1972)}]{GS}%
  \BibitemOpen
  \bibfield  {author} {\bibinfo {author} {\bibfnamefont {R.~W.}\ \bibnamefont
  {Gerchberg}},\ }\href {https://api.semanticscholar.org/CorpusID:55691159}
  {\bibfield  {journal} {\bibinfo  {journal} {Optik}\ }\textbf {\bibinfo
  {volume} {35}},\ \bibinfo {pages} {237} (\bibinfo {year} {1972})}\BibitemShut
  {NoStop}%
\bibitem [{\citenamefont {Zuo}\ \emph {et~al.}(2020)\citenamefont {Zuo},
  \citenamefont {Li}, \citenamefont {Sun}, \citenamefont {Fan}, \citenamefont
  {Zhang}, \citenamefont {Lu}, \citenamefont {Zhang}, \citenamefont {Wang},
  \citenamefont {Huang},\ and\ \citenamefont {Chen}}]{zuo2020transport}%
  \BibitemOpen
  \bibfield  {author} {\bibinfo {author} {\bibfnamefont {C.}~\bibnamefont
  {Zuo}}, \bibinfo {author} {\bibfnamefont {J.}~\bibnamefont {Li}}, \bibinfo
  {author} {\bibfnamefont {J.}~\bibnamefont {Sun}}, \bibinfo {author}
  {\bibfnamefont {Y.}~\bibnamefont {Fan}}, \bibinfo {author} {\bibfnamefont
  {J.}~\bibnamefont {Zhang}}, \bibinfo {author} {\bibfnamefont
  {L.}~\bibnamefont {Lu}}, \bibinfo {author} {\bibfnamefont {R.}~\bibnamefont
  {Zhang}}, \bibinfo {author} {\bibfnamefont {B.}~\bibnamefont {Wang}},
  \bibinfo {author} {\bibfnamefont {L.}~\bibnamefont {Huang}},\ and\ \bibinfo
  {author} {\bibfnamefont {Q.}~\bibnamefont {Chen}},\ }\href@noop {} {\bibfield
   {journal} {\bibinfo  {journal} {Optics and Lasers in Engineering}\ }\textbf
  {\bibinfo {volume} {135}},\ \bibinfo {pages} {106187} (\bibinfo {year}
  {2020})}\BibitemShut {NoStop}%
\bibitem [{\citenamefont {Primot}\ and\ \citenamefont
  {Sogno}(1995)}]{primot1995achromatic}%
  \BibitemOpen
  \bibfield  {author} {\bibinfo {author} {\bibfnamefont {J.}~\bibnamefont
  {Primot}}\ and\ \bibinfo {author} {\bibfnamefont {L.}~\bibnamefont {Sogno}},\
  }\href@noop {} {\bibfield  {journal} {\bibinfo  {journal} {JOSA A}\ }\textbf
  {\bibinfo {volume} {12}},\ \bibinfo {pages} {2679} (\bibinfo {year}
  {1995})}\BibitemShut {NoStop}%
\bibitem [{\citenamefont {Platt}\ and\ \citenamefont
  {Shack}(2001)}]{SH-history}%
  \BibitemOpen
  \bibfield  {author} {\bibinfo {author} {\bibfnamefont {B.~C.}\ \bibnamefont
  {Platt}}\ and\ \bibinfo {author} {\bibfnamefont {R.}~\bibnamefont {Shack}},\
  }\href@noop {} {\bibinfo {title} {History and principles of shack-hartmann
  wavefront sensing}} (\bibinfo {year} {2001})\BibitemShut {NoStop}%
\bibitem [{\citenamefont {Dai}\ \emph {et~al.}(2016{\natexlab{a}})\citenamefont
  {Dai}, \citenamefont {Li}, \citenamefont {Wang},\ and\ \citenamefont
  {Bu}}]{DAI2016264}%
  \BibitemOpen
  \bibfield  {author} {\bibinfo {author} {\bibfnamefont {F.}~\bibnamefont
  {Dai}}, \bibinfo {author} {\bibfnamefont {J.}~\bibnamefont {Li}}, \bibinfo
  {author} {\bibfnamefont {X.}~\bibnamefont {Wang}},\ and\ \bibinfo {author}
  {\bibfnamefont {Y.}~\bibnamefont {Bu}},\ }\href
  {https://doi.org/https://doi.org/10.1016/j.optcom.2016.01.068} {\bibfield
  {journal} {\bibinfo  {journal} {Optics Communications}\ }\textbf {\bibinfo
  {volume} {367}},\ \bibinfo {pages} {264} (\bibinfo {year}
  {2016}{\natexlab{a}})}\BibitemShut {NoStop}%
\bibitem [{\citenamefont {Pathak}\ and\ \citenamefont
  {Boruah}(2014)}]{Pathak_2014}%
  \BibitemOpen
  \bibfield  {author} {\bibinfo {author} {\bibfnamefont {B.}~\bibnamefont
  {Pathak}}\ and\ \bibinfo {author} {\bibfnamefont {B.~R.}\ \bibnamefont
  {Boruah}},\ }\href {https://doi.org/10.1088/2040-8978/16/5/055403} {\bibfield
   {journal} {\bibinfo  {journal} {Journal of Optics}\ }\textbf {\bibinfo
  {volume} {16}},\ \bibinfo {pages} {055403} (\bibinfo {year}
  {2014})}\BibitemShut {NoStop}%
\bibitem [{\citenamefont {Southwell}(1980)}]{southwell1980wave}%
  \BibitemOpen
  \bibfield  {author} {\bibinfo {author} {\bibfnamefont {W.~H.}\ \bibnamefont
  {Southwell}},\ }\href@noop {} {\bibfield  {journal} {\bibinfo  {journal}
  {JOsA}\ }\textbf {\bibinfo {volume} {70}},\ \bibinfo {pages} {998} (\bibinfo
  {year} {1980})}\BibitemShut {NoStop}%
\bibitem [{\citenamefont {Chanteloup}(2005)}]{chanteloup2005multiple}%
  \BibitemOpen
  \bibfield  {author} {\bibinfo {author} {\bibfnamefont {J.-C.}\ \bibnamefont
  {Chanteloup}},\ }\href@noop {} {\bibfield  {journal} {\bibinfo  {journal}
  {Applied optics}\ }\textbf {\bibinfo {volume} {44}},\ \bibinfo {pages} {1559}
  (\bibinfo {year} {2005})}\BibitemShut {NoStop}%
\bibitem [{\citenamefont {Tian}\ \emph {et~al.}(1995)\citenamefont {Tian},
  \citenamefont {Itoh},\ and\ \citenamefont {Yatagai}}]{tian1995simple}%
  \BibitemOpen
  \bibfield  {author} {\bibinfo {author} {\bibfnamefont {X.}~\bibnamefont
  {Tian}}, \bibinfo {author} {\bibfnamefont {M.}~\bibnamefont {Itoh}},\ and\
  \bibinfo {author} {\bibfnamefont {T.}~\bibnamefont {Yatagai}},\ }\href@noop
  {} {\bibfield  {journal} {\bibinfo  {journal} {Applied optics}\ }\textbf
  {\bibinfo {volume} {34}},\ \bibinfo {pages} {7213} (\bibinfo {year}
  {1995})}\BibitemShut {NoStop}%
\bibitem [{\citenamefont {Liang}\ \emph {et~al.}(1994)\citenamefont {Liang},
  \citenamefont {Grimm}, \citenamefont {Goelz},\ and\ \citenamefont
  {Bille}}]{liang1994objective}%
  \BibitemOpen
  \bibfield  {author} {\bibinfo {author} {\bibfnamefont {J.}~\bibnamefont
  {Liang}}, \bibinfo {author} {\bibfnamefont {B.}~\bibnamefont {Grimm}},
  \bibinfo {author} {\bibfnamefont {S.}~\bibnamefont {Goelz}},\ and\ \bibinfo
  {author} {\bibfnamefont {J.~F.}\ \bibnamefont {Bille}},\ }\href@noop {}
  {\bibfield  {journal} {\bibinfo  {journal} {JOSA A}\ }\textbf {\bibinfo
  {volume} {11}},\ \bibinfo {pages} {1949} (\bibinfo {year}
  {1994})}\BibitemShut {NoStop}%
\bibitem [{\citenamefont {He}\ \emph {et~al.}(2021)\citenamefont {He},
  \citenamefont {Liu}, \citenamefont {Ning}, \citenamefont {Li}, \citenamefont
  {Xu},\ and\ \citenamefont {Jiang}}]{He:21}%
  \BibitemOpen
  \bibfield  {author} {\bibinfo {author} {\bibfnamefont {Y.}~\bibnamefont
  {He}}, \bibinfo {author} {\bibfnamefont {Z.}~\bibnamefont {Liu}}, \bibinfo
  {author} {\bibfnamefont {Y.}~\bibnamefont {Ning}}, \bibinfo {author}
  {\bibfnamefont {J.}~\bibnamefont {Li}}, \bibinfo {author} {\bibfnamefont
  {X.}~\bibnamefont {Xu}},\ and\ \bibinfo {author} {\bibfnamefont
  {Z.}~\bibnamefont {Jiang}},\ }\href {https://doi.org/10.1364/OE.427261}
  {\bibfield  {journal} {\bibinfo  {journal} {Opt. Express}\ }\textbf {\bibinfo
  {volume} {29}},\ \bibinfo {pages} {17669} (\bibinfo {year}
  {2021})}\BibitemShut {NoStop}%
\bibitem [{\citenamefont {Seifert}\ \emph {et~al.}(2005)\citenamefont
  {Seifert}, \citenamefont {Tiziani},\ and\ \citenamefont
  {Osten}}]{seifert2005wavefront}%
  \BibitemOpen
  \bibfield  {author} {\bibinfo {author} {\bibfnamefont {L.}~\bibnamefont
  {Seifert}}, \bibinfo {author} {\bibfnamefont {H.}~\bibnamefont {Tiziani}},\
  and\ \bibinfo {author} {\bibfnamefont {W.}~\bibnamefont {Osten}},\
  }\href@noop {} {\bibfield  {journal} {\bibinfo  {journal} {Optics
  communications}\ }\textbf {\bibinfo {volume} {245}},\ \bibinfo {pages} {255}
  (\bibinfo {year} {2005})}\BibitemShut {NoStop}%
\bibitem [{\citenamefont {Howard}\ \emph {et~al.}(2023)\citenamefont {Howard},
  \citenamefont {Esslinger}, \citenamefont {Wang}, \citenamefont {Norreys},\
  and\ \citenamefont {D{\"o}pp}}]{howard2023hyperspectral}%
  \BibitemOpen
  \bibfield  {author} {\bibinfo {author} {\bibfnamefont {S.}~\bibnamefont
  {Howard}}, \bibinfo {author} {\bibfnamefont {J.}~\bibnamefont {Esslinger}},
  \bibinfo {author} {\bibfnamefont {R.~H.}\ \bibnamefont {Wang}}, \bibinfo
  {author} {\bibfnamefont {P.}~\bibnamefont {Norreys}},\ and\ \bibinfo {author}
  {\bibfnamefont {A.}~\bibnamefont {D{\"o}pp}},\ }\href@noop {} {\bibfield
  {journal} {\bibinfo  {journal} {High Power Laser Science and Engineering}\
  }\textbf {\bibinfo {volume} {11}},\ \bibinfo {pages} {e32} (\bibinfo {year}
  {2023})}\BibitemShut {NoStop}%
\bibitem [{\citenamefont {Harbers}\ \emph {et~al.}(1996)\citenamefont
  {Harbers}, \citenamefont {Kunst},\ and\ \citenamefont
  {Leibbrandt}}]{harbers1996analysis}%
  \BibitemOpen
  \bibfield  {author} {\bibinfo {author} {\bibfnamefont {G.}~\bibnamefont
  {Harbers}}, \bibinfo {author} {\bibfnamefont {P.}~\bibnamefont {Kunst}},\
  and\ \bibinfo {author} {\bibfnamefont {G.}~\bibnamefont {Leibbrandt}},\
  }\href@noop {} {\bibfield  {journal} {\bibinfo  {journal} {Applied optics}\
  }\textbf {\bibinfo {volume} {35}},\ \bibinfo {pages} {6162} (\bibinfo {year}
  {1996})}\BibitemShut {NoStop}%
\bibitem [{\citenamefont {Dai}\ \emph {et~al.}(2016{\natexlab{b}})\citenamefont
  {Dai}, \citenamefont {Zheng}, \citenamefont {Bu},\ and\ \citenamefont
  {Wang}}]{dai2016modal}%
  \BibitemOpen
  \bibfield  {author} {\bibinfo {author} {\bibfnamefont {F.}~\bibnamefont
  {Dai}}, \bibinfo {author} {\bibfnamefont {Y.}~\bibnamefont {Zheng}}, \bibinfo
  {author} {\bibfnamefont {Y.}~\bibnamefont {Bu}},\ and\ \bibinfo {author}
  {\bibfnamefont {X.}~\bibnamefont {Wang}},\ }\href@noop {} {\bibfield
  {journal} {\bibinfo  {journal} {Applied Optics}\ }\textbf {\bibinfo {volume}
  {56}},\ \bibinfo {pages} {61} (\bibinfo {year}
  {2016}{\natexlab{b}})}\BibitemShut {NoStop}%
\bibitem [{\citenamefont {Shen}\ \emph {et~al.}(2019)\citenamefont {Shen},
  \citenamefont {Wang}, \citenamefont {Xie}, \citenamefont {Min}, \citenamefont
  {Fu}, \citenamefont {Liu}, \citenamefont {Gong},\ and\ \citenamefont
  {Yuan}}]{vortex}%
  \BibitemOpen
  \bibfield  {author} {\bibinfo {author} {\bibfnamefont {Y.}~\bibnamefont
  {Shen}}, \bibinfo {author} {\bibfnamefont {X.}~\bibnamefont {Wang}}, \bibinfo
  {author} {\bibfnamefont {Z.}~\bibnamefont {Xie}}, \bibinfo {author}
  {\bibfnamefont {C.}~\bibnamefont {Min}}, \bibinfo {author} {\bibfnamefont
  {X.}~\bibnamefont {Fu}}, \bibinfo {author} {\bibfnamefont {Q.}~\bibnamefont
  {Liu}}, \bibinfo {author} {\bibfnamefont {M.}~\bibnamefont {Gong}},\ and\
  \bibinfo {author} {\bibfnamefont {X.}~\bibnamefont {Yuan}},\ }\href@noop {}
  {\bibfield  {journal} {\bibinfo  {journal} {Light: Science \& Applications}\
  }\textbf {\bibinfo {volume} {8}},\ \bibinfo {pages} {90} (\bibinfo {year}
  {2019})}\BibitemShut {NoStop}%
\bibitem [{\citenamefont {McGloin}\ and\ \citenamefont
  {Dholakia}(2005)}]{mcgloin2005bessel}%
  \BibitemOpen
  \bibfield  {author} {\bibinfo {author} {\bibfnamefont {D.}~\bibnamefont
  {McGloin}}\ and\ \bibinfo {author} {\bibfnamefont {K.}~\bibnamefont
  {Dholakia}},\ }\href@noop {} {\bibfield  {journal} {\bibinfo  {journal}
  {Contemporary physics}\ }\textbf {\bibinfo {volume} {46}},\ \bibinfo {pages}
  {15} (\bibinfo {year} {2005})}\BibitemShut {NoStop}%
\bibitem [{\citenamefont {Efremidis}\ \emph {et~al.}(2019)\citenamefont
  {Efremidis}, \citenamefont {Chen}, \citenamefont {Segev},\ and\ \citenamefont
  {Christodoulides}}]{efremidis2019airy}%
  \BibitemOpen
  \bibfield  {author} {\bibinfo {author} {\bibfnamefont {N.~K.}\ \bibnamefont
  {Efremidis}}, \bibinfo {author} {\bibfnamefont {Z.}~\bibnamefont {Chen}},
  \bibinfo {author} {\bibfnamefont {M.}~\bibnamefont {Segev}},\ and\ \bibinfo
  {author} {\bibfnamefont {D.~N.}\ \bibnamefont {Christodoulides}},\
  }\href@noop {} {\bibfield  {journal} {\bibinfo  {journal} {Optica}\ }\textbf
  {\bibinfo {volume} {6}},\ \bibinfo {pages} {686} (\bibinfo {year}
  {2019})}\BibitemShut {NoStop}%
\bibitem [{\citenamefont {Noll}(1976)}]{noll1976zernike}%
  \BibitemOpen
  \bibfield  {author} {\bibinfo {author} {\bibfnamefont {R.~J.}\ \bibnamefont
  {Noll}},\ }\href@noop {} {\bibfield  {journal} {\bibinfo  {journal} {JOsA}\
  }\textbf {\bibinfo {volume} {66}},\ \bibinfo {pages} {207} (\bibinfo {year}
  {1976})}\BibitemShut {NoStop}%
\bibitem [{\citenamefont {Tyson}\ and\ \citenamefont
  {Frazier}(2022)}]{tyson2022principles}%
  \BibitemOpen
  \bibfield  {author} {\bibinfo {author} {\bibfnamefont {R.~K.}\ \bibnamefont
  {Tyson}}\ and\ \bibinfo {author} {\bibfnamefont {B.~W.}\ \bibnamefont
  {Frazier}},\ }\href@noop {} {\emph {\bibinfo {title} {Principles of adaptive
  optics}}}\ (\bibinfo  {publisher} {CRC press},\ \bibinfo {year}
  {2022})\BibitemShut {NoStop}%
\bibitem [{\citenamefont {Andersen}(2018)}]{andersen2018efficient}%
  \BibitemOpen
  \bibfield  {author} {\bibinfo {author} {\bibfnamefont {T.~B.}\ \bibnamefont
  {Andersen}},\ }\href@noop {} {\bibfield  {journal} {\bibinfo  {journal}
  {Optics Express}\ }\textbf {\bibinfo {volume} {26}},\ \bibinfo {pages}
  {18878} (\bibinfo {year} {2018})}\BibitemShut {NoStop}%
\bibitem [{\citenamefont {Chen}\ \emph {et~al.}(2001)\citenamefont {Chen},
  \citenamefont {Donoho},\ and\ \citenamefont {Saunders}}]{chen2001atomic}%
  \BibitemOpen
  \bibfield  {author} {\bibinfo {author} {\bibfnamefont {S.~S.}\ \bibnamefont
  {Chen}}, \bibinfo {author} {\bibfnamefont {D.~L.}\ \bibnamefont {Donoho}},\
  and\ \bibinfo {author} {\bibfnamefont {M.~A.}\ \bibnamefont {Saunders}},\
  }\href@noop {} {\bibfield  {journal} {\bibinfo  {journal} {SIAM review}\
  }\textbf {\bibinfo {volume} {43}},\ \bibinfo {pages} {129} (\bibinfo {year}
  {2001})}\BibitemShut {NoStop}%
\bibitem [{\citenamefont {Rubinstein}\ \emph {et~al.}(2010)\citenamefont
  {Rubinstein}, \citenamefont {Bruckstein},\ and\ \citenamefont
  {Elad}}]{rubinstein2010dictionaries}%
  \BibitemOpen
  \bibfield  {author} {\bibinfo {author} {\bibfnamefont {R.}~\bibnamefont
  {Rubinstein}}, \bibinfo {author} {\bibfnamefont {A.~M.}\ \bibnamefont
  {Bruckstein}},\ and\ \bibinfo {author} {\bibfnamefont {M.}~\bibnamefont
  {Elad}},\ }\href@noop {} {\bibfield  {journal} {\bibinfo  {journal}
  {Proceedings of the IEEE}\ }\textbf {\bibinfo {volume} {98}},\ \bibinfo
  {pages} {1045} (\bibinfo {year} {2010})}\BibitemShut {NoStop}%
\bibitem [{\citenamefont {Cand{\`e}s}\ \emph {et~al.}(2006)\citenamefont
  {Cand{\`e}s}, \citenamefont {Romberg},\ and\ \citenamefont
  {Tao}}]{candes2006robust}%
  \BibitemOpen
  \bibfield  {author} {\bibinfo {author} {\bibfnamefont {E.~J.}\ \bibnamefont
  {Cand{\`e}s}}, \bibinfo {author} {\bibfnamefont {J.}~\bibnamefont
  {Romberg}},\ and\ \bibinfo {author} {\bibfnamefont {T.}~\bibnamefont {Tao}},\
  }\href@noop {} {\bibfield  {journal} {\bibinfo  {journal} {IEEE Transactions
  on information theory}\ }\textbf {\bibinfo {volume} {52}},\ \bibinfo {pages}
  {489} (\bibinfo {year} {2006})}\BibitemShut {NoStop}%
\bibitem [{\citenamefont {Candes}\ and\ \citenamefont
  {Tao}(2006)}]{candes2006near}%
  \BibitemOpen
  \bibfield  {author} {\bibinfo {author} {\bibfnamefont {E.~J.}\ \bibnamefont
  {Candes}}\ and\ \bibinfo {author} {\bibfnamefont {T.}~\bibnamefont {Tao}},\
  }\href@noop {} {\bibfield  {journal} {\bibinfo  {journal} {IEEE transactions
  on information theory}\ }\textbf {\bibinfo {volume} {52}},\ \bibinfo {pages}
  {5406} (\bibinfo {year} {2006})}\BibitemShut {NoStop}%
\bibitem [{\citenamefont {Jaderberg}\ \emph {et~al.}(2015)\citenamefont
  {Jaderberg}, \citenamefont {Simonyan}, \citenamefont {Zisserman} \emph
  {et~al.}}]{jaderberg2015spatial}%
  \BibitemOpen
  \bibfield  {author} {\bibinfo {author} {\bibfnamefont {M.}~\bibnamefont
  {Jaderberg}}, \bibinfo {author} {\bibfnamefont {K.}~\bibnamefont {Simonyan}},
  \bibinfo {author} {\bibfnamefont {A.}~\bibnamefont {Zisserman}}, \emph
  {et~al.},\ }\href@noop {} {\bibfield  {journal} {\bibinfo  {journal}
  {Advances in neural information processing systems}\ }\textbf {\bibinfo
  {volume} {28}} (\bibinfo {year} {2015})}\BibitemShut {NoStop}%
\bibitem [{\citenamefont {Beck}\ and\ \citenamefont
  {Teboulle}(2009)}]{beck2009fast}%
  \BibitemOpen
  \bibfield  {author} {\bibinfo {author} {\bibfnamefont {A.}~\bibnamefont
  {Beck}}\ and\ \bibinfo {author} {\bibfnamefont {M.}~\bibnamefont
  {Teboulle}},\ }\href@noop {} {\bibfield  {journal} {\bibinfo  {journal} {SIAM
  journal on imaging sciences}\ }\textbf {\bibinfo {volume} {2}},\ \bibinfo
  {pages} {183} (\bibinfo {year} {2009})}\BibitemShut {NoStop}%
\bibitem [{\citenamefont {Kim}\ \emph {et~al.}(2007)\citenamefont {Kim},
  \citenamefont {Koh}, \citenamefont {Lustig}, \citenamefont {Boyd},\ and\
  \citenamefont {Gorinevsky}}]{kim2007interior}%
  \BibitemOpen
  \bibfield  {author} {\bibinfo {author} {\bibfnamefont {S.-J.}\ \bibnamefont
  {Kim}}, \bibinfo {author} {\bibfnamefont {K.}~\bibnamefont {Koh}}, \bibinfo
  {author} {\bibfnamefont {M.}~\bibnamefont {Lustig}}, \bibinfo {author}
  {\bibfnamefont {S.}~\bibnamefont {Boyd}},\ and\ \bibinfo {author}
  {\bibfnamefont {D.}~\bibnamefont {Gorinevsky}},\ }\href@noop {} {\bibfield
  {journal} {\bibinfo  {journal} {IEEE journal of selected topics in signal
  processing}\ }\textbf {\bibinfo {volume} {1}},\ \bibinfo {pages} {606}
  (\bibinfo {year} {2007})}\BibitemShut {NoStop}%
\bibitem [{\citenamefont {Boyd}\ \emph {et~al.}(2011)\citenamefont {Boyd},
  \citenamefont {Parikh}, \citenamefont {Chu}, \citenamefont {Peleato},
  \citenamefont {Eckstein} \emph {et~al.}}]{boyd2011distributed}%
  \BibitemOpen
  \bibfield  {author} {\bibinfo {author} {\bibfnamefont {S.}~\bibnamefont
  {Boyd}}, \bibinfo {author} {\bibfnamefont {N.}~\bibnamefont {Parikh}},
  \bibinfo {author} {\bibfnamefont {E.}~\bibnamefont {Chu}}, \bibinfo {author}
  {\bibfnamefont {B.}~\bibnamefont {Peleato}}, \bibinfo {author} {\bibfnamefont
  {J.}~\bibnamefont {Eckstein}}, \emph {et~al.},\ }\href@noop {} {\bibfield
  {journal} {\bibinfo  {journal} {Foundations and Trend in Machine learning}\
  }\textbf {\bibinfo {volume} {3}},\ \bibinfo {pages} {1} (\bibinfo {year}
  {2011})}\BibitemShut {NoStop}%
\bibitem [{\citenamefont {Kingma}(2014)}]{kingma2014adam}%
  \BibitemOpen
  \bibfield  {author} {\bibinfo {author} {\bibfnamefont {D.~P.}\ \bibnamefont
  {Kingma}},\ }\href@noop {} {\bibfield  {journal} {\bibinfo  {journal} {arXiv
  preprint arXiv:1412.6980}\ } (\bibinfo {year} {2014})}\BibitemShut {NoStop}%
\bibitem [{\citenamefont {McLeod}(1954)}]{axicon}%
  \BibitemOpen
  \bibfield  {author} {\bibinfo {author} {\bibfnamefont {J.~H.}\ \bibnamefont
  {McLeod}},\ }\href@noop {} {\bibfield  {journal} {\bibinfo  {journal} {JOSA}\
  }\textbf {\bibinfo {volume} {44}},\ \bibinfo {pages} {592} (\bibinfo {year}
  {1954})}\BibitemShut {NoStop}%
\bibitem [{\citenamefont {Garc{\'e}s-Ch{\'a}vez}\ \emph
  {et~al.}(2002)\citenamefont {Garc{\'e}s-Ch{\'a}vez}, \citenamefont {McGloin},
  \citenamefont {Melville}, \citenamefont {Sibbett},\ and\ \citenamefont
  {Dholakia}}]{garces2002simultaneous}%
  \BibitemOpen
  \bibfield  {author} {\bibinfo {author} {\bibfnamefont {V.}~\bibnamefont
  {Garc{\'e}s-Ch{\'a}vez}}, \bibinfo {author} {\bibfnamefont {D.}~\bibnamefont
  {McGloin}}, \bibinfo {author} {\bibfnamefont {H.}~\bibnamefont {Melville}},
  \bibinfo {author} {\bibfnamefont {W.}~\bibnamefont {Sibbett}},\ and\ \bibinfo
  {author} {\bibfnamefont {K.}~\bibnamefont {Dholakia}},\ }\href@noop {}
  {\bibfield  {journal} {\bibinfo  {journal} {Nature}\ }\textbf {\bibinfo
  {volume} {419}},\ \bibinfo {pages} {145} (\bibinfo {year}
  {2002})}\BibitemShut {NoStop}%
\bibitem [{\citenamefont {Pedregosa}\ \emph {et~al.}(2011)\citenamefont
  {Pedregosa}, \citenamefont {Varoquaux}, \citenamefont {Gramfort},
  \citenamefont {Michel}, \citenamefont {Thirion}, \citenamefont {Grisel},
  \citenamefont {Blondel}, \citenamefont {Prettenhofer}, \citenamefont {Weiss},
  \citenamefont {Dubourg}, \citenamefont {Vanderplas}, \citenamefont {Passos},
  \citenamefont {Cournapeau}, \citenamefont {Brucher}, \citenamefont {Perrot},\
  and\ \citenamefont {Duchesnay}}]{scikit-learn}%
  \BibitemOpen
  \bibfield  {author} {\bibinfo {author} {\bibfnamefont {F.}~\bibnamefont
  {Pedregosa}}, \bibinfo {author} {\bibfnamefont {G.}~\bibnamefont
  {Varoquaux}}, \bibinfo {author} {\bibfnamefont {A.}~\bibnamefont {Gramfort}},
  \bibinfo {author} {\bibfnamefont {V.}~\bibnamefont {Michel}}, \bibinfo
  {author} {\bibfnamefont {B.}~\bibnamefont {Thirion}}, \bibinfo {author}
  {\bibfnamefont {O.}~\bibnamefont {Grisel}}, \bibinfo {author} {\bibfnamefont
  {M.}~\bibnamefont {Blondel}}, \bibinfo {author} {\bibfnamefont
  {P.}~\bibnamefont {Prettenhofer}}, \bibinfo {author} {\bibfnamefont
  {R.}~\bibnamefont {Weiss}}, \bibinfo {author} {\bibfnamefont
  {V.}~\bibnamefont {Dubourg}}, \bibinfo {author} {\bibfnamefont
  {J.}~\bibnamefont {Vanderplas}}, \bibinfo {author} {\bibfnamefont
  {A.}~\bibnamefont {Passos}}, \bibinfo {author} {\bibfnamefont
  {D.}~\bibnamefont {Cournapeau}}, \bibinfo {author} {\bibfnamefont
  {M.}~\bibnamefont {Brucher}}, \bibinfo {author} {\bibfnamefont
  {M.}~\bibnamefont {Perrot}},\ and\ \bibinfo {author} {\bibfnamefont
  {E.}~\bibnamefont {Duchesnay}},\ }\href@noop {} {\bibfield  {journal}
  {\bibinfo  {journal} {Journal of Machine Learning Research}\ }\textbf
  {\bibinfo {volume} {12}},\ \bibinfo {pages} {2825} (\bibinfo {year}
  {2011})}\BibitemShut {NoStop}%
\bibitem [{\citenamefont {Tibshirani}(1996)}]{lasso}%
  \BibitemOpen
  \bibfield  {author} {\bibinfo {author} {\bibfnamefont {R.}~\bibnamefont
  {Tibshirani}},\ }\href@noop {} {\bibfield  {journal} {\bibinfo  {journal}
  {Journal of the Royal Statistical Society Series B: Statistical Methodology}\
  }\textbf {\bibinfo {volume} {58}},\ \bibinfo {pages} {267} (\bibinfo {year}
  {1996})}\BibitemShut {NoStop}%
\bibitem [{\citenamefont {Lopez-Quintas}\ \emph {et~al.}(2020)\citenamefont
  {Lopez-Quintas}, \citenamefont {Holgado}, \citenamefont {Drevinskas},
  \citenamefont {Kazansky}, \citenamefont {Sola},\ and\ \citenamefont
  {Alonso}}]{swaveplate}%
  \BibitemOpen
  \bibfield  {author} {\bibinfo {author} {\bibfnamefont {I.}~\bibnamefont
  {Lopez-Quintas}}, \bibinfo {author} {\bibfnamefont {W.}~\bibnamefont
  {Holgado}}, \bibinfo {author} {\bibfnamefont {R.}~\bibnamefont {Drevinskas}},
  \bibinfo {author} {\bibfnamefont {P.~G.}\ \bibnamefont {Kazansky}}, \bibinfo
  {author} {\bibfnamefont {{\'I}.~J.}\ \bibnamefont {Sola}},\ and\ \bibinfo
  {author} {\bibfnamefont {B.}~\bibnamefont {Alonso}},\ }\href@noop {}
  {\bibfield  {journal} {\bibinfo  {journal} {Journal of Optics}\ }\textbf
  {\bibinfo {volume} {22}},\ \bibinfo {pages} {095402} (\bibinfo {year}
  {2020})}\BibitemShut {NoStop}%
\bibitem [{\citenamefont {L{\'o}pez-Ripa}\ \emph {et~al.}(2022)\citenamefont
  {L{\'o}pez-Ripa}, \citenamefont {Sola},\ and\ \citenamefont
  {Alonso}}]{lopez2022bulk}%
  \BibitemOpen
  \bibfield  {author} {\bibinfo {author} {\bibfnamefont {M.}~\bibnamefont
  {L{\'o}pez-Ripa}}, \bibinfo {author} {\bibfnamefont {{\'I}.~J.}\ \bibnamefont
  {Sola}},\ and\ \bibinfo {author} {\bibfnamefont {B.}~\bibnamefont {Alonso}},\
  }\href@noop {} {\bibfield  {journal} {\bibinfo  {journal} {Photonics
  Research}\ }\textbf {\bibinfo {volume} {10}},\ \bibinfo {pages} {922}
  (\bibinfo {year} {2022})}\BibitemShut {NoStop}%
\bibitem [{\citenamefont {Weisse}\ \emph {et~al.}(2023)\citenamefont {Weisse},
  \citenamefont {Esslinger}, \citenamefont {Howard}, \citenamefont {Foerster},
  \citenamefont {Haberstroh}, \citenamefont {Doyle}, \citenamefont {Norreys},
  \citenamefont {Schreiber}, \citenamefont {Karsch},\ and\ \citenamefont
  {D{\"o}pp}}]{nils}%
  \BibitemOpen
  \bibfield  {author} {\bibinfo {author} {\bibfnamefont {N.}~\bibnamefont
  {Weisse}}, \bibinfo {author} {\bibfnamefont {J.}~\bibnamefont {Esslinger}},
  \bibinfo {author} {\bibfnamefont {S.}~\bibnamefont {Howard}}, \bibinfo
  {author} {\bibfnamefont {F.}~\bibnamefont {Foerster}}, \bibinfo {author}
  {\bibfnamefont {F.}~\bibnamefont {Haberstroh}}, \bibinfo {author}
  {\bibfnamefont {L.}~\bibnamefont {Doyle}}, \bibinfo {author} {\bibfnamefont
  {P.}~\bibnamefont {Norreys}}, \bibinfo {author} {\bibfnamefont
  {J.}~\bibnamefont {Schreiber}}, \bibinfo {author} {\bibfnamefont
  {S.}~\bibnamefont {Karsch}},\ and\ \bibinfo {author} {\bibfnamefont
  {A.}~\bibnamefont {D{\"o}pp}},\ }\href@noop {} {\bibfield  {journal}
  {\bibinfo  {journal} {Optics Express}\ }\textbf {\bibinfo {volume} {31}},\
  \bibinfo {pages} {19733} (\bibinfo {year} {2023})}\BibitemShut {NoStop}%
\bibitem [{\citenamefont {Alonso}\ \emph {et~al.}(2020)\citenamefont {Alonso},
  \citenamefont {Holgado},\ and\ \citenamefont {Sola}}]{alonso2020compact}%
  \BibitemOpen
  \bibfield  {author} {\bibinfo {author} {\bibfnamefont {B.}~\bibnamefont
  {Alonso}}, \bibinfo {author} {\bibfnamefont {W.}~\bibnamefont {Holgado}},\
  and\ \bibinfo {author} {\bibfnamefont {{\'I}.~J.}\ \bibnamefont {Sola}},\
  }\href@noop {} {\bibfield  {journal} {\bibinfo  {journal} {Optics Express}\
  }\textbf {\bibinfo {volume} {28}},\ \bibinfo {pages} {15625} (\bibinfo {year}
  {2020})}\BibitemShut {NoStop}%
\bibitem [{\citenamefont {L{\'o}pez-Ripa}\ \emph {et~al.}(2024)\citenamefont
  {L{\'o}pez-Ripa}, \citenamefont {P{\'e}rez-Benito}, \citenamefont {Alonso},
  \citenamefont {Weigand},\ and\ \citenamefont {Sola}}]{lopez2024few}%
  \BibitemOpen
  \bibfield  {author} {\bibinfo {author} {\bibfnamefont {M.}~\bibnamefont
  {L{\'o}pez-Ripa}}, \bibinfo {author} {\bibfnamefont {{\'O}.}~\bibnamefont
  {P{\'e}rez-Benito}}, \bibinfo {author} {\bibfnamefont {B.}~\bibnamefont
  {Alonso}}, \bibinfo {author} {\bibfnamefont {R.}~\bibnamefont {Weigand}},\
  and\ \bibinfo {author} {\bibfnamefont {{\'I}.}~\bibnamefont {Sola}},\
  }\href@noop {} {\bibfield  {journal} {\bibinfo  {journal} {Optics Express}\
  }\textbf {\bibinfo {volume} {32}},\ \bibinfo {pages} {21149} (\bibinfo {year}
  {2024})}\BibitemShut {NoStop}%
\bibitem [{\citenamefont {Tibshirani}(2022)}]{tibshirani2022coordinate}%
  \BibitemOpen
  \bibfield  {author} {\bibinfo {author} {\bibfnamefont {R.}~\bibnamefont
  {Tibshirani}},\ }\href@noop {} {\bibfield  {journal} {\bibinfo  {journal}
  {Course Convex Optimization}\ ,\ \bibinfo {pages} {10}} (\bibinfo {year}
  {2022})}\BibitemShut {NoStop}%
\end{thebibliography}%

\appendix
\section{Representation of Vortex in Zernike Basis}

We aim to expand the optical vortex phase function \(\phi(\theta) = \ell \theta\) in terms of Zernike polynomials \(Z_n^m(\rho, \theta)\). The Zernike polynomials form a complete set over the unit disk, allowing any function to be represented as:

\[
\phi(\rho, \theta) = \sum_{n=0}^{\infty} \sum_{m=-n}^{n} c_n^m Z_n^m(\rho, \theta)
\]

Due to the orthogonality of the Zernike polynomials, the coefficients \(c_n^m\) are given by:

\[
c_n^m = \frac{\int_{0}^{1} \int_{0}^{2\pi} \phi(\theta) Z_n^{m*}(\rho, \theta) \rho\, d\theta\, d\rho}{\int_{0}^{1} \int_{0}^{2\pi} |Z_n^m(\rho, \theta)|^2 \rho\, d\theta\, d\rho}
\]

Since \(\phi(\theta)\) is independent of \(\rho\), we can separate the radial and angular components. The denominator simplifies to \(\pi\) due to normalization, and we define the radial integral:

\[
S_n^{|m|} = \int_{0}^{1} R_n^{|m|}(\rho) \rho\, d\rho
\]

where \(R_n^{|m|}(\rho)\) are the radial Zernike polynomials. Our main task is then to evaluate the angular integral:

\[
A_m = \int_{0}^{2\pi} \phi(\theta) e^{-i m \theta}\, d\theta = \ell \int_{0}^{2\pi} \theta e^{-i m \theta}\, d\theta
\]

This integral can be evaluated using integration by parts. Let \(u = \theta\) and \(dv = e^{-i m \theta} d\theta\), so \(du = d\theta\) and \(v = \frac{e^{-i m \theta}}{-i m}\). Applying integration by parts:
\begin{widetext}
\[
A_m = u v \Big|_{0}^{2\pi} - \int_{0}^{2\pi} v\, du = \left[ \theta \cdot \frac{e^{-i m \theta}}{-i m} \right]_{0}^{2\pi} - \int_{0}^{2\pi} \frac{e^{-i m \theta}}{-i m}\, d\theta = \frac{2\pi}{-i m}
\]
\end{widetext}

Therefore, the coefficient \(c_n^m\) becomes:

\[
c_n^m = \frac{\ell\, S_n^{|m|}\, A_m}{\pi} = \frac{\ell\, S_n^{|m|}\, \left( \frac{2\pi}{-i m} \right)}{\pi} = \ell\, \frac{2 i}{m}\, S_n^{|m|}
\]

This shows that the coefficients \(c_n^m\) decay proportionally to \(1/m\). The slow decay of \(c_n^m\) with increasing \(m\) indicates that a large number of Zernike modes are required to accurately represent the vortex phase \(\phi(\theta) = \ell \theta\). Therefore, the Zernike basis is inefficient for representing optical vortices, motivating the use of alternative representations such as an over-complete dictionary.

\section{LASSO Regression}

The least absolute shrinkage and selection operator (LASSO) is a technique that leverages the L1 norm to find regularized solutions to least squares problems. Consider one has a measurement vector, $\vec{y}$, and a dictionary, $\mathbf{x}$. One wishes to find the sparsest possible coefficient vector $c$, subject to the condition that $\vec{y} = \mathbf{x}\vec{c}$. The minimization problem described by LASSO is,

\begin{equation}
    \underset{c}{\text{min}}\left\{\|\vec{y} - \mathbf{x}\vec{c}\|_{2} 
 + \alpha \|\vec{c}\|_{1} \right\}
\end{equation}
where $||a||_2 = \sum_{j}a^2_{j}$, $||c||_1 = \sum_{j}|c_{j}|$, and $\alpha$ is a constant determining the level of sparsity. 

The scikit-learn implementation solves this problem using coordinate descent. Here, the modal coefficients are optimized one at a time\cite{tibshirani2022coordinate}. To optimize parameter $c_i$, we first define $\mathbf{x}_{-i}$ and $c_{-i}$ as the dictionary and coefficients with index $i$ removed, meaning the remaining residual for component $i$ is found by $(y - \mathbf{x}_{-i}c_{-i})$. Solving the LASSO equation for $c_i$ gives,

\begin{equation}
    c_i = \frac{\mathbf{x}^{T}_{i}(y - \mathbf{x}_{-i}c_{-i})}{\mathbf{x}_{i}^{T}\mathbf{x}_{i}} - \frac{\alpha}{\mathbf{x}_{i}^{T}\mathbf{x}_{i}} \frac{\partial |c_{i}|}{\partial c_{i}}.
\end{equation}

where $\frac{\partial |c_i|}{\partial c_i} = \text{sign}(c_i)$. This equation clearly has a term from the least squares, and a term from the L1 regularization. To enforce sparsity, a soft threshold is used to set the parameter to zero if its value is under a threshold.

\begin{equation}
    c_i = \text{sign}(c_{i})\cdot\text{max}\left(\left|\frac{\mathbf{x}^{T}_{i}(y - \mathbf{x}_{-i}c_{-i})}{\mathbf{x}_{i}^{T}\mathbf{x}_{i}}\right|  - \frac{\alpha}{\mathbf{x}_{i}^{T}\mathbf{x}_{i}}, 0\right)
\end{equation}

This process is iterated over all coefficients $i \in [1,N_c]$, and is then repeated until the final error falls below a predefined threshold, or a predefined termination number of loops is reached. 

\end{document}